\documentclass[prl,preprint,showpacs,floatfix]{revtex4}
\usepackage{graphicx}
\usepackage{color}

\def\revision{}

\begin{document}
\title{Element-resolved thermodynamics of magnetocaloric LaFe$_{13-x}$Si$_x$}

\author{M. E. Gruner$^{1,2}$\footnote{Corresponding author: Markus.Gruner@uni-due.de}}
\author{W. Keune$^{1,3}$}
\author{B. Roldan Cuenya$^{4}$}
\author{C. Weis$^{1}$}
\author{J. Landers$^{1}$}
\author{S. I. Makarov$^{1,3}$}
\author{D. Klar$^{1}$}
\author{M. Y. Hu$^{5}$}
\author{E. E. Alp$^{5}$}
\author{J. Zhao$^{5}$}
\author{M. Krautz$^{2}$}
\author{O. Gutfleisch$^{6}$}
\author{H. Wende$^{1}$}
\affiliation{$^{1}$University of Duisburg-Essen and Center for
  Nanointegration Duisburg-Essen (CENIDE), D-47048 Duisburg, Germany}
\affiliation{$^{2}$IFW Dresden P.O. Box 270116, 01171 Dresden, Germany}
\affiliation{$^{3}$Max Planck Institute of Microstructure Physics, 06120 Halle, Germany}
\affiliation{$^{4}$Department of Physics, Ruhr-University Bochum, 44780 Bochum, Germany}
\affiliation{$^{5}$Advanced Photon Source, Argonne National Laboratory, Argonne, IL 60439, United States}
\affiliation{$^{6}$Materials Science, TU Darmstadt, 64287 Darmstadt, Germany}

\def\figurewidth{0.5\textwidth}

\begin{abstract}
By combination of two independent approaches, 
nuclear resonant inelastic X-ray scattering and first-principles
calculations in the framework of density functional theory,
{\revision
we demonstrate significant changes in the
element-resolved vibrational density of states 
across the first-order transition from the ferromagnetic
low temperature to the paramagnetic high temperature phase of
LaFe$_{13-x}$Si$_x$. These changes originate from the
itinerant electron metamagnetism associated with Fe
and lead to a pronounced magneto\-elastic softening despite the
large
volume decrease at the transition.
The increase in lattice entropy associated with the Fe subsystem
is significant and contributes cooperatively with the magnetic and electronic
entropy changes to the excellent magneto- and barocaloric
properties.}
\end{abstract}


\maketitle
Ferroic materials allow for
significant adiabatic temperature changes
induced by realistic electrical and magnetic fields, by
external stress and under pressure
\cite{cn:Tishin03,cn:Gschneidner05,cn:Gutfleisch11AM,cn:Faehler12FerroicCooling}.
This allows their use in
solid state refrigeration concepts as an energy efficient
alternative to the classical gas-compressor scheme.
A good cooling material is characterized by
a large isothermal entropy change $|\Delta S_{\rm iso}|$, which
determines the latent heat to be taken up during a first-order
transformation in conjunction with a high adiabatic temperature
change $|\Delta T_{\rm ad}|$.
Apart from the prototype Gd-based systems \cite{cn:Pecharsky97PRL},
a large number of suitable materials
were proposed, which undergo a magnetic first-order transition
and perform well in both respects
(e.\,g., \cite{cn:Sandeman12Scripta}).
Among the outstanding materials are LaFe$_{13-x}$Si$_x$-based systems
(1.0$\,\leq\,$$x$$\,\leq\,$1.6)
\cite{cn:Fujieda02APL,cn:Lyubina10AM,preparation}, which
consist of largely abundant components 
\cite{cn:Gutfleisch11AM}.
Their structure corresponds to
cubic NaZn$_{13}$ ($Fm\overline{3}c$, 112 atoms in the unit cell),
with two distinguished crystallographic Fe-sites, Fe$_{\rm I}$ and
Fe$_{\rm II}$,
on the 8-fold (8b) and 96-fold (96i) Wyckoff-positions, respectively.
La resides on (8a)-sites, while Si shares the (96i)
site with Fe$_{\rm II}$ \cite{cn:Chang03JoPD,cn:Hamdeh04,cn:Rosca10JALCOM}.
The Curie temperature, $T_{\rm C}$, of the ferro- (FM) to paramagnetic (PM)
transformation is around
$200\,$K, depending on composition.
$T_{\rm C}$ increases proportionally with increasing Si-content 
\cite{cn:Palstra83,cn:Jia06JAP},
but the transition changes to second-order, while $|\Delta S_{\rm
  iso}|$ and $|\Delta T_{\rm ad}|$ decrease significantly.
By concomitant hydrogenation and Mn substitution, $T_{\rm C}$ can be precisely
adjusted to ambient conditions
without severe degradation of the caloric performance
\cite{cn:Fujita03H,cn:Wang03CP,cn:Barcza11IEEE,cn:Krautz14JALCOM}.
The first-order magnetic transformation is accompanied by an abrupt
isostructural
volume decrease of $1\%$ for $x$$\,=\,$1.5 upon the loss of magnetic order
\cite{cn:Hu01APL}.
This also gives rise to a large  {\em inverse}
barocaloric effect \cite{cn:Manosa11NCOMMS}.
In the FM phase the thermal expansion coefficient is largely
reduced or even negative \cite{cn:Chang03JoPD,cn:Huang13JACS},
which presents similarities with the Invar-type
thermal expansion anomalies
discovered in Fe$_{65}$Ni$_{35}$ (and other ferrous alloys) more than
one hundred years ago (e.\,g., \cite{cn:Wassermann90}).

Consequently, the moment-volume-instability of La-Fe-Si has been
discussed in terms of the {\em itinerant electron metamagnetism} (IEM)
arising from the (partially) non-localized character of the
Fe moments \cite{cn:Fujita99,cn:Fujita01} within
a phenomenological Ginzburg-Landau description
\cite{cn:Fujita03,cn:Yamada03}.
The IEM picture gained further support from
first-principles calculations
\cite{cn:Kuzmin07,cn:Fujita12Scripta,cn:Gercsi14} through the
identification of metastable minima of the binding surface,
which correspond to metastable magnetic configurations
at distinct volumina.
This is also a characteristic feature of Fe-Ni Invar, where
the compensation of thermal expansion
is linked to the redistribution of charge between non-bonding majority
spin-states above and anti-bonding minority spin states below the
Fermi-level \cite{cn:Entel93}.
However, further {\em ab initio} work on La-Fe-Si
concentrates on the electronic structure in the FM
phase and the non-spinpolarized state
\cite{cn:Fujita03,cn:Wang06JMMM,cn:Han08,cn:Boutahar13}, while
thorough characterization of the paramagnetic phase is still missing.

In this letter we will establish the link between
the electronic structure of La-Fe-Si and its
macroscopic thermodynamic behavior
in both, the FM and the PM phase,
by a combination of nuclear resonant inelastic X-ray
scattering (NRIXS) and {\em ab initio}
lattice dynamics.
{\revision
We demonstrate for the first time that temperature-induced
magnetic disorder causes distinct modifications in the
vibrational density of states (VDOS)
of a cubic metal.}
We disentangle the elemental contributions
to the
VDOS, which determines
the intrinsic vibrational thermodynamic properties,
such as lattice entropy,
and relate them to phase-induced changes in the electronic structure.

The isothermal entropy change is usually divided up as
$\Delta S_{\rm iso}=\Delta S_{\rm mag}+\Delta S_{\rm lat}+\Delta S_{\rm el}$,
i.\,e., into the contributions from the magnetic, lattice and
electronic degrees of freedom, respectively \cite{cn:Tishin03}.
These are associated with the
configuration entropy arising from spin disorder, excitation of
quasi-harmonic phonons and thermal occupation changes of the electronic
states, respectively.
Mixed interactions are often not taken into account separately,
although they might
enhance $\Delta S_{\rm iso}$ significantly \cite{cn:Mukherjee11PRB}
and can be expected for $3d$ metals \cite{cn:Tishin03}.
Anharmonic contributions are neglected here as well.
For La-Fe-Si, $\Delta S_{\rm mag}$ is expected to be the
driving contribution, while $\Delta S_{\rm el}$ is considered
negligible and $\Delta S_{\rm lat}$ is deemed to counteract
assuming a quasiharmonic renormalization of the Debye
temperature $\Theta$ \cite{cn:Ranke05,cn:Jia06JAP}.
In the following we will demonstrate, that both,
$\Delta S_{\rm lat}$ and $\Delta S_{\rm el}$ 
contribute {\em cooperatively} to the magnetocaloric effect.
The sign of $\Delta S_{\rm lat}$ is determined by the
IEM of Fe, which affects magneto\-elastic interactions and
thus modifies phonon frequencies oppositely to what is expected from
Gr\"uneisen theory.

For the measurements, {\revision
we used
a polycrystalline sample with nominal composition
LaFe$_{11.6}$Si$_{1.4}$ (Fe with 10\% enrichment in $^{57}$Fe);
for details, see \cite{supplementary}.}
The sample was characterized by X-ray diffraction
and 14.4-keV M\"ossbauer backscattering spectroscopy
\cite{supplementary}, showing that
89\,$\pm$1\,\% of the Fe atoms
in the sample are in the La(Fe,Si)$_{13}$ phase.
11\,$\pm$1\,\%
are in the bcc Fe secondary phase
\cite{preparation,cn:Krautz14JALCOM}.
Magnetization measurements reveal (in agreement with
\cite{cn:Fujieda02APL}) a first-order FM-PM transition at
$T_{\rm C}$$\,=\,$189\,K with a hysteresis of 3\,K.
NRIXS \cite{cn:Seto95PRL,cn:Sturhahn95PRL,cn:Chumakov95EPL}
was performed at Sector-3 beamline at the Advanced
Photon Source, Argonne National Laboratory.  The incident X-ray energy
was around $E_0$$\,=\,$14.412\,keV, the nuclear resonance energy of $^{57}$Fe.
The X-ray beam is highly monochromatized with an energy bandwidth of
1\,meV \cite{cn:Toellner00}.
NRIXS was carried out in zero external magnetic field
at four different measurement temperatures $T_{\rm exp}$,
two in the FM phase (62\,K, 164\,K) and two in the PM phase
(220\,K, 299\,K).
The $^{57}$Fe-specific VDOS
were extracted from the NRIXS data
using the PHOENIX program \cite{cn:Sturhahn00} and
corrected for the $\alpha$-Fe contribution \cite{supplementary}.

\begin{figure}
\includegraphics[width=\figurewidth]{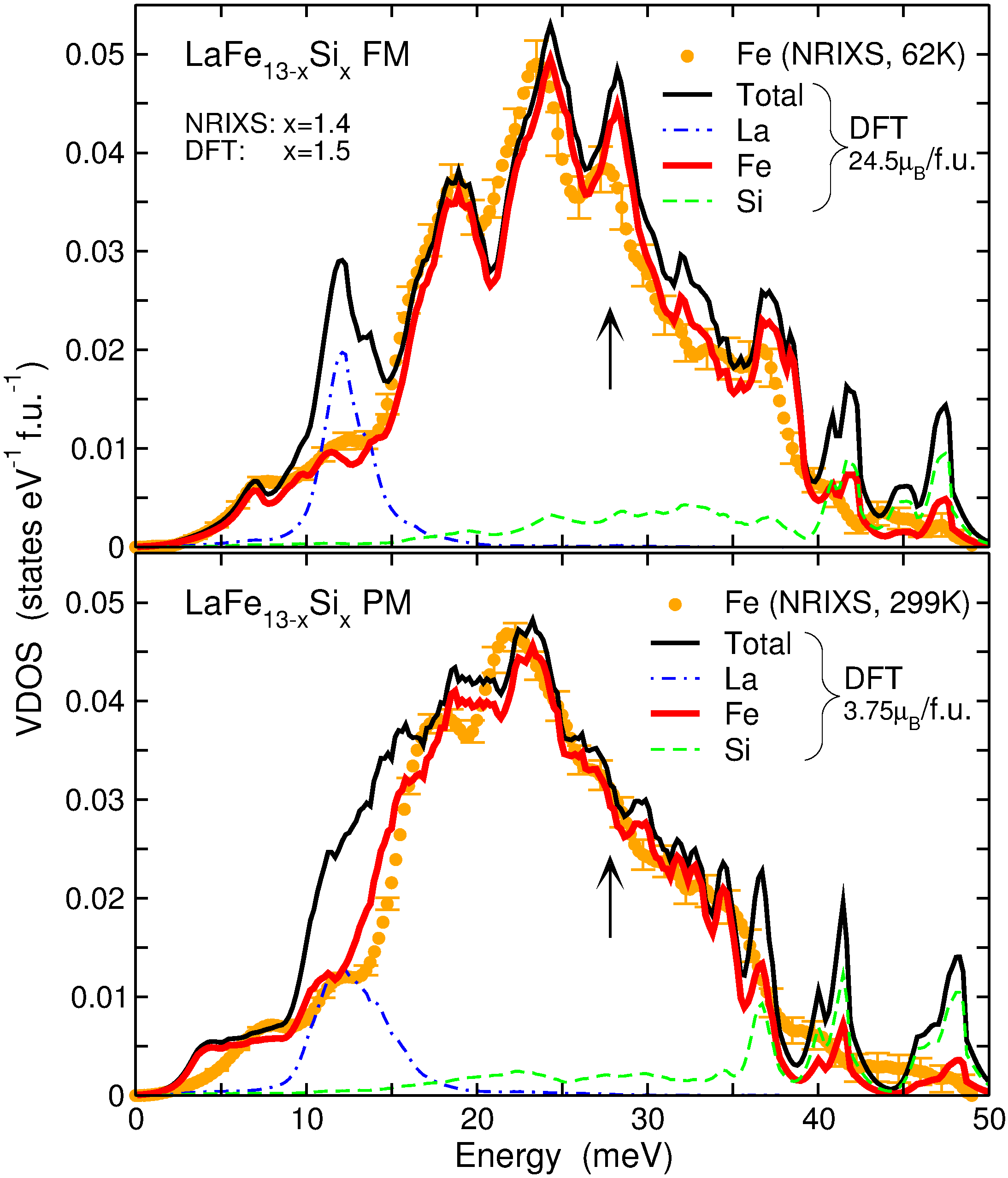}
\caption{(Color online) Element-resolved
VDOS of FM and PM LaFe$_{1-x}$Si$_{x}$.
Orange circles with error bars denote the partial $^{57}$Fe-DOS obtained with
NRIXS at $T_{\rm exp}$$\,=\,$$62\,$K and
$T_{\rm exp}$$\,=\,$$299\,$K, respectively,
corrected for the bcc Fe secondary phase contribution \cite{supplementary}.
The lines refer to the DFT results for different magnetizations
($M$$\,=\,$$24.5\mu_{\rm B}$/f.u. and $M$$\,=\,$$3.75\mu_{\rm
  B}$/f.u., respectively). The thick red lines denote the partial DOS
of the Fe atoms, which compares to the NRIXS measurement. The thinner
black, the green dotted and the blue dash dotted lines refer to the
total VDOS and the partial contributions of Si and La, respectively.
}\label{fig:PDOS}
\end{figure}

The {\em ab initio} part is carried out 
with the VASP package \cite{cn:VASP1,cn:VASP2}
in the framework of density functional theory (DFT)
using the generalized gradient approximation (GGA) \cite{cn:Perdew91}.
We represented the 112 atom unit cell
by a 28 atom primitive cell with fcc basis and
introduced three Si atoms on the (96i) sites, i.\,e.,
$x$$\,=\,$1.5\,Si per formula unit (f.u.), 
such that the space group is
minimally reduced to rhombohedral ($R3$)
with still $12$ inequivalent lattice sites.
For the PM phase, we carefully determined a stable
collinear spin arrangement with 10 inverted Fe moments, which has
a total spin-magnetization of 3.75$\,\mu_{\rm B}$/f.u.\ compared to
24.5$\,\mu_{\rm B}$/f.u.\ for FM. An
average spin-moment of 1.7$\,\mu_{\rm B}$/Fe was obtained for PM
as compared
to 2.2$\,\mu_{\rm B}$/Fe for FM
in good agreement with available
M\"ossbauer, neutron diffraction and DFT data
\cite{cn:Palstra83,cn:Wang03JPCM,cn:Rosca10JALCOM,cn:Fujita12Scripta}.
We optimized ionic positions and volume before
the dynamical matrix was constructed with Dario Alf\`e's PHON code
\cite{cn:Alfe09PHON} based on the Hellmann-Feynman forces obtained from
56 $\pm$-displacements in a
$2$$\times$$2$$\times$$2$ supercell. This yields the
VDOS, $g(E)$, from which we obtain
thermodynamic quantities
like the lattice entropy $S_{\rm lat}$ and their
temperature dependence \cite{cn:Grimvall86,cn:Fultz10}.
We restrict to the harmonic
approximation using the equilibrium volume of both magnetic
states since
thermal expansion is small or absent below and immediately above $T_{\rm  C}$.
For further details, see
\cite{supplementary}.

\begin{figure}
\centering
\includegraphics[width=\figurewidth]{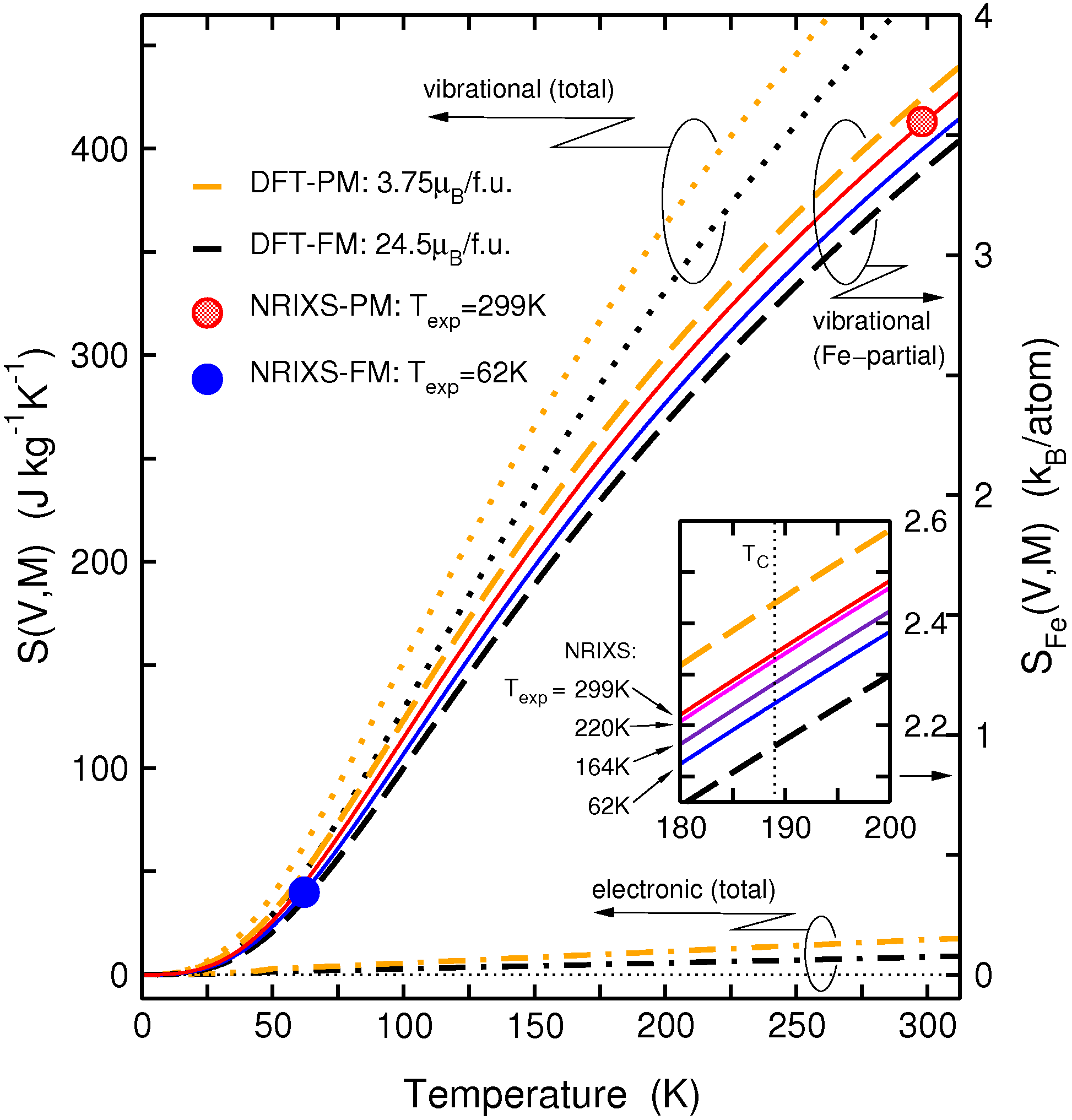}
\caption{(Color online) Total (left scale)
and Fe-contribution (right scale and inset) to
  the lattice entropy and electronic entropy of the FM and PM phases.
 For a Si content of $x$$\,=\,$1.5 (DFT) both scales are equivalent.
  The lines are calculated from $g(E)$ for the volume $V$ and
  magnetization $M$
  corresponding to the measurement temperature $T_{\rm exp}$
  (circles) or $T$$\,=\,$0 (DFT), respectively.
}\label{fig:Sv}
\end{figure}

{\revision
We observe striking differences in the experimental Fe-projected VDOS below
and above T$_{\rm C}$ 
(Fig. \ref{fig:PDOS}), in particular the disappearance of the distinct phonon
peak near 28 meV (arrows) in the PM state.
This unusual 17\%-effect (see $\alpha$-Fe as reference \cite{cn:Neuhaus14PRB})
is a distinct manifestation of the impact of magnetic disorder on both,
the low- and high-energy part of the phonon spectra
which is known from iron oxide \cite{cn:Struzhkin01PRL} but has not been demonstrated
in this clarity for a metallic system.
DFT reproduces all relevant features of the partial VDOS of Fe obtained with NRIXS,
including the uniform shift to lower energies above $T_{\rm C}$, which}
verifies our La and Si partial and
the total VDOS, which we can only obtain from DFT.
Thus, we encounter an overall softening in $g(E)$ upon heating, which
overrides the stiffening expected from Gr\"uneisen theory,
according to the volume contraction.
From the NRIXS $g(E)$ we
obtain the temperature dependent lattice entropy $S_{\rm lat}(M_{\rm exp},V_{\rm exp})$
corresponding to volume $V$ and magnetization $M$ at the respective
$T_{\rm exp}$ using the well-known textbook relation \cite{cn:Fultz10}
{\revision
neglecting the T-dependence of $M$.
Fig.\ \ref{fig:Sv} shows that
changing from FM to PM configurations results
in an increase in $S_{\rm lat}$ at $T_{\rm C}$,}
which amounts to
$\Delta S_{\rm lat}^{\rm
  Fe}|_{T_{\rm C}}$$\,=\,$$11\,$J\,kg$^{-1}$K$^{-1}$$\,=\,$$0.10\,k_{\rm B}$/Fe
calculated from the NRIXS VDOS for $T_{\rm exp}$$\,=\,$62\,K and 299\,K.
This is about one half of
$\Delta S_{\rm iso}$$\,=\,$$24\,$J\,kg$^{-1}$K$^{-1}$
from literature \cite{cn:Fujieda02APL}, 
{\revision
obtained from integrating specific heat
across the field-induced transition.}
From our NRIXS data closer to $T_{\rm C}$
 ($T_{\rm exp}$$\,=\,$164\,K and 220\,K)
we obtain a reduced 
$\Delta S_{\rm lat}^{\rm
  Fe}|_{T_{\rm C}}$$\,=\,$$5\,$J\,kg$^{-1}$K$^{-1}$.
As thermal expansion is largely canceled, the difference of $6\,$J\,kg$^{-1}$K$^{-1}$
originates from the increasing spin disorder in the FM phase and the remaining spin
correlation in the PM phase and is therefore another manifestation of the strong
magnetoelastic coupling in La-Fe-Si, 
arising from the observed changes in the VDOS, $g(E)$, across the phase transition.  
The entropy Debye temperature, $\Theta_{\rm S}$, derived 
from the logarithmic moment of the partial Fe NRIXS $g(E)$ 
\cite{cn:Rosen83PRB,cn:Grimvall86,cn:Eriksson92PRB},
decreases by 3\,\% from $\Theta^{\rm
  Fe}_{\rm 62\,K}$$\,=\,$$371\,$K (FM) to
$\Theta^{\rm Fe}_{\rm 299\,K}$$\,=\,$$360\,$K (PM),
whereas normal Gr\"uneisen behavior would rather result in a 1-2\,\% increase
 due to the large negative volume change at $T_{\rm C}$.
The DFT model fully confirms this trend and yields an even larger
$\Delta S_{\rm lat}^{\rm
  tot}|_{T_{\rm C}}=32\,$J\,kg$^{-1}$K$^{-1}$  from the total $g(E)$,
which coincides with the computed partial Fe-contribution.
\begin{figure}
\centering
\includegraphics[width=\figurewidth]{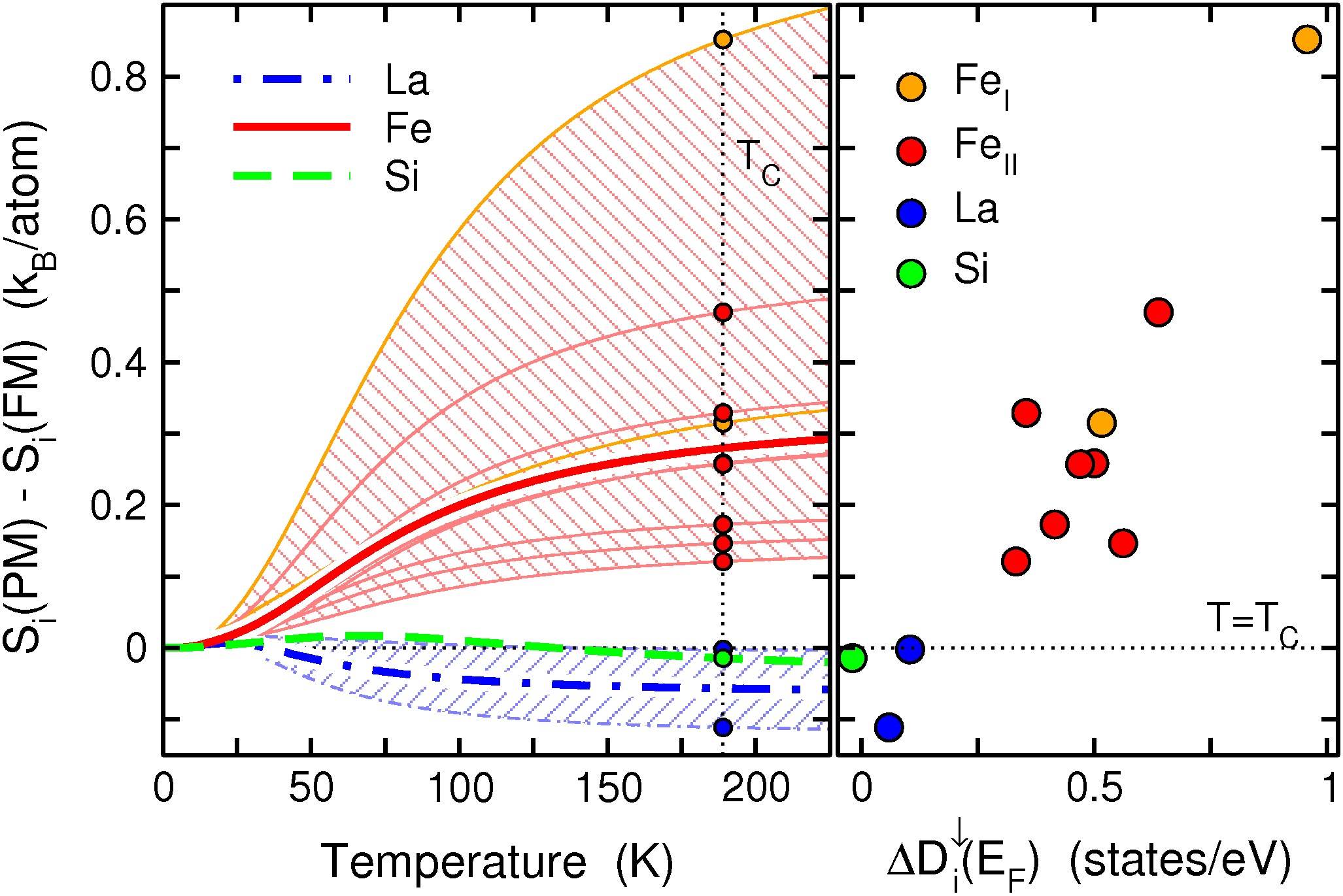}
\caption{(Color online) Element- and site-resolved difference
  in lattice entropy $\Delta S_{{\rm lat},i}$$\,=\,$$S_{{\rm lat},i}({\rm
    PM})-S_{{\rm lat},i}({\rm FM})$ from DFT (left side).
  The thick lines refer to the elemental averages,
  thin lines refer to values for the inequivalent lattice
  sites $i$. The right graph demonstrates
  the correlation between the site-resolved
  $\Delta S_{{\rm lat},i}$ at $T_{\rm C}$
  and the change in the site-resolved minority spin
  DOS at $E_{\rm F}$,
  $\Delta D_i^{\downarrow}(E_{\rm F})$$\,=\,$$
  D_i^{\downarrow}(E_{\rm F},{\rm PM})-D_i^{\downarrow}(E_{\rm F},{\rm
    FM})$.
}\label{fig:DiffSv}
\end{figure}
\begin{figure}
\centering
\includegraphics[width=\figurewidth]{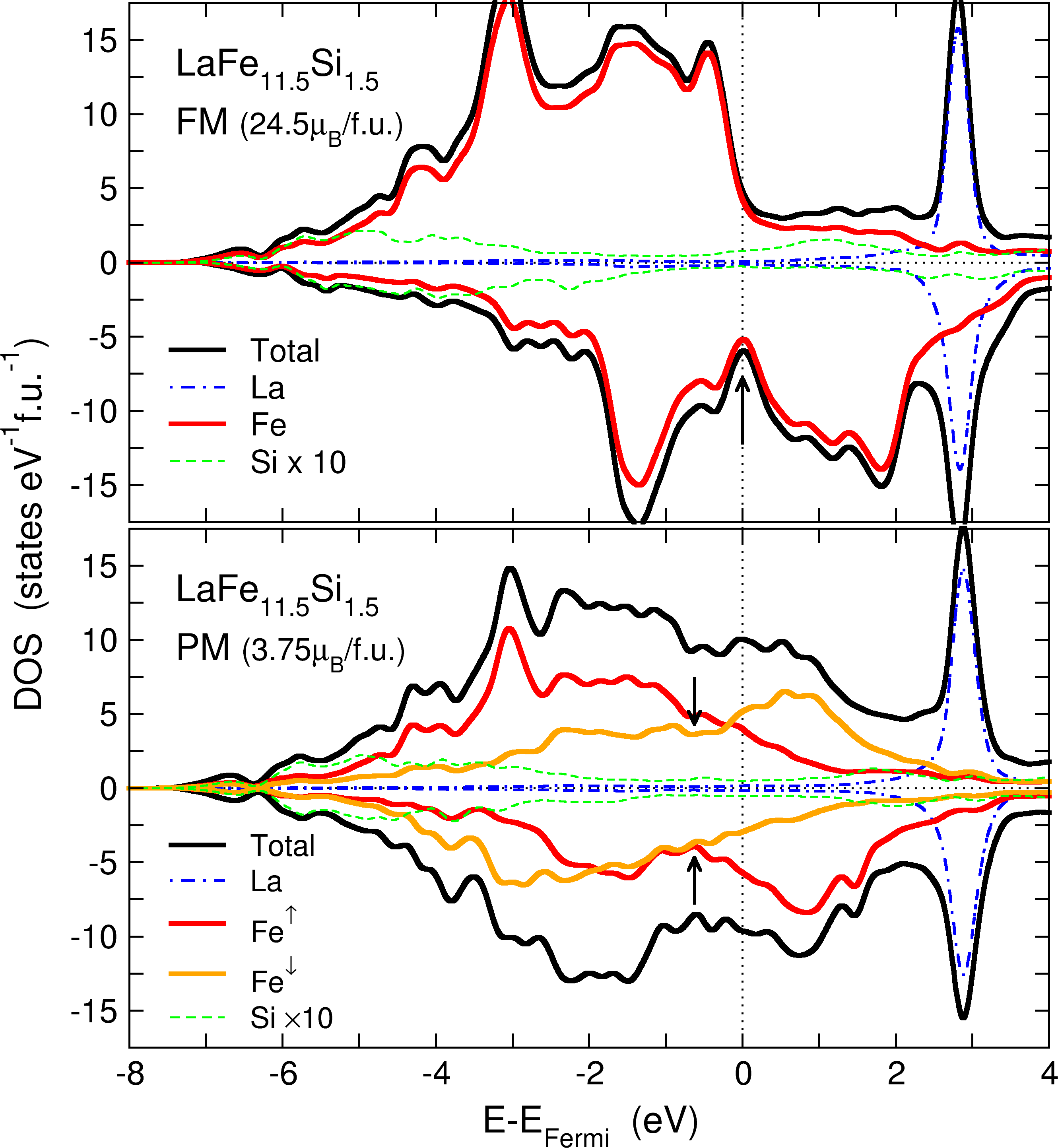}
\caption{(Color online) Total and element-resolved
  electronic DOS of ferromagnetic
($M$$\,=\,$$24.5\mu_{\rm B}$/f.u., top)
and paramagnetic ($M$$\,=\,$$3.75\mu_{\rm B}$/f.u., bottom)
LaFe$_{11.5}$Si$_{1.5}$ from DFT. The majority spin channel is denoted
by positive values, the minority channel by negative.
}\label{fig:EDOS}
\end{figure}
{\revision
The elemental decomposition of $\Delta S_{\rm lat}$ in Fig.\ \ref{fig:DiffSv} (left) 
shows some spread according to the
chemical and electronic configuration, which have also a strong influence
on the interatomic spacings \cite{cn:Liu03JPCM,cn:Wang03JPCM,cn:Rosca10JALCOM}, in particular
for the Fe sites, which encounter a change in magnetic order.
Here, the average contribution (thick line) is large and positive, while 
for La and Si it is vanishing or even slightly negative.
Since symmetry does not change, the anomalous sign of
$\Delta S_{\rm lat}$ is thus solely related to the change of the
magnetic environment of the Fe atoms.

This trend can be explained through adiabatic electron phonon coupling, which 
has recently been identified as the cause of anomalous softening
or stiffening in several Fe-based materials
\cite{cn:Delaire08PRL,cn:Delaire11PNAS,cn:Munoz11PRL}
upon temperature-dependent positional and chemical disorder. Such disorder
broadens minima or maxima in the electronic density of states (DOS), $D(E)$,
around $E_{\rm F}$, where a high availability of electronic states assists the
screening of perturbations from a displaced atom.
Indeed, we can identify a likewise correlation
between PM and FM phase in the site-resolved lattice entropy changes  
$\Delta S_{{\rm lat},i}$ and
in the site-resolved DOS $\Delta D_i(E_{\rm F})$ of the respective site $i$.
This trend is particularly pronounced for the minority 
electrons corresponding the respective site (Fig.\ \ref{fig:DiffSv}, right)
and originates from the IEM of Fe, as shown by the
electronic DOS.
The FM DOS in the upper panel of Fig.\ \ref{fig:EDOS},
exhibits a completely filled majority $d$-channel and a nearly
half-filled minority channel \cite{cn:Fujita03,cn:Wang06JMMM,cn:Han08,cn:Boutahar13}.
The large exchange splitting moves the mid-$d$-band minimum found
in the majority channel at $-2\,$eV right to the Fermi level $E_{\rm
  F}$ in the minority channel, which is a stabilizing feature for the FM phase.
La and Si states are essentially absent in this important energy range.
As typical for an IEM, changing magnetic order (lower panel) distorts the minority
Fe-DOS (Fe$^{\uparrow}$ and Fe$^{\downarrow}$, for both magnetization directions, respectively)
as states hybridize with the majority channel of neighboring antiparallel Fe.
The local magnetic moment, i.\,e., the
exchange splitting at each site decreases, which shifts the remainders of 
the minimum away from $E_{\rm F}$ (arrows). 
This suggests that magnetic disorder in La-Fe-Si affects restoring forces in a similar
fashion as chemical ordering in FeV
\cite{cn:Munoz11PRL}. 
This also changes the}
Sommerfeld constant for the
electronic specific heat 
{\revision
by a factor of two
($\gamma_{\rm PM}$$\,=\,$$56.1\,$mJ\,kg$^{-1}$K$^{-2}$
vs.\ $\gamma_{\rm FM}$$\,=\,$$28.8\,$mJ\,kg$^{-1}$K$^{-2}$),
leading to the} 
cooperative contribution of the
electronic subsystem to the phase transition
(cf.\ Fig.\ \ref{fig:Sv}), similar to metamagnetic
$\alpha$-FeRh \cite{cn:Deak14}.

{\revision
We conclude that in La-Fe-Si magnetic disorder causes
unique changes in the VDOS.
The consequence are significant
cooperative contributions of magnetism, lattice and electrons
to the entropy change, which
provide the foundation for the
excellent magneto- and barocaloric properties of this compound.
The electronic DOS minimum at $E_{\rm F}$ in the FM phase in combination with
the itinerant nature of Fe-magnetism is
responsible for the anomalous magneto\-elastic softening
and the magnitude of $\Delta S_{\rm lat}$ and $\Delta S_{\rm el}$.
Both favor low-volume, low-moment configurations 
contributing to the Invar-type (over-)compensation of
thermal expansion in the FM phase and foster an early,
first-order-type transformation to the magnetically disordered phase.
However, since our results indicate a strong interaction of all relevant
 degrees of freedom (i.\,e.,
electronic, vibrational and magnetic) the
common decomposition of $\Delta S_{\rm iso}$ into three independent
entropy terms must be interpreted with caution.

La-Fe-Si thus provides an ideal model system to unravel the
contributions to magnetocaloric and Invar effect.
Combining large-scale first-principles calculations 
and state-of-the-art scattering techniques provides
the essential step to identify the microscopic mechanisms.
As for other ferrous systems, NRIXS has proven an ideal
experimental method to determine the
specific vibrational contribution to the entropy change.
In turn, 
we see the approximate modelling of structural and magnetic disorder
in the 28 atom pseudo-ordered primitive cell, which grants us access to the electronic scale,
justified by the excellent agreement in the
VDOS of the Fe atoms. This approach provides
thus a suitable basis for the exploration of improved materials and compositions.
Our work suggests that maximizing the Fe-content is the primary strategy
to improve the magnetocaloric performance of the material, if the band-filling 
is adjusted carefully by additional components.}

\begin{acknowledgments}
The authors would like to thank P.\ Entel (Duisburg-Essen),
G.\ Bayreuther and J.\ Kirschner (Halle) and S.\ F\"ahler (Dresden) for
important discussions and support. We are grateful to
U.\ v.\ H\"orsten (Duisburg-Essen) 
and Wenli Bi (Argonne) for technical assistance.
Calculations were carried out on the
massively parallel computers (Cray XT6/m and Opterox)
of the Center of Computational Sciences and Simulation, CCSS, of the
University of Duisburg-Essen.
Funding by the DFG via SPP1239, SPP1599 and SPP1538 is gratefully
acknowledged. BRC (RUB/UCF) was funded by the US National Science
Foundation (NSF-DMR 1207065).
Use of the Advanced Photon Source, an Office of Science User Facility
operated for the U.S.\ Department of Energy (DOE) Office of Science by
Argonne National Laboratory, was supported by the U.S.\ DOE
(DE-AC02-06CH11357).
\end{acknowledgments}


\end{document}